\documentstyle[11pt,newpasp,twoside,psfig]{article}
\begin{document}

\title{ISO Observations of Pre-Main Sequence and Vega-type Stars}
\author{M.E. van den Ancker}
\affil{Harvard-Smithsonian Center for Astrophysics, 60 Garden Street, 
MS 42, Cambridge, MA 02138, USA}

\begin{abstract}
I present an overview of the results obtained by the {\it Infrared 
Space Observatory} (ISO) on circumstellar material in pre-main 
sequence (PMS) stars. Results obtained for embedded YSOs, Herbig 
Ae/Be systems and T~Tauri stars are reviewed and their connection to 
the disks around Vega-type systems is discussed. Although the gas 
contents of the PMS environment will also be discussed briefly, this 
review will mainly focus on the composition, mineralogy 
and evolution of dust in these systems, and the results will be 
compared to those found in other classes of objects, including solar 
system comets.
\end{abstract}

\section{Introduction}
One of the most exciting developments in the study of pre-main 
sequence and young main sequence stars is the advent of space-based 
infrared observatories. The first infrared space mission, IRAS, 
discovered the presence of excess infrared emission above 
photospheric levels in many young stellar objects (YSOs) due 
to the presence of circumstellar disks or envelopes containing 
small dust particles. More surprisingly, IRAS also discovered 
infrared excesses in several nearby main sequence stars. These 
Vega-type stars (after the prototype Vega; Aumann et al. 1984)
have infrared excesses that usually start at longer 
wavelengths (typically $>$ 25~$\mu$m) than in YSOs, indicative 
of much cooler dust temperatures. A particularly intriguing case 
is the A-type star $\beta$~Pictoris, exhibiting both an infrared 
excess and a circumstellar disk seen in scattered light, 
which may form the evolutionary link between YSOs and 
Vega-type stars. 
Thirteen years after the IRAS mission its successor saw first 
light: the {\it Infrared Space Observatory}\footnote{Based on 
observations with ISO, an ESA project with instruments
funded by ESA Member States (especially the PI countries: France,
Germany, the Netherlands and the United Kingdom) and with the
participation of ISAS and NASA.} (ISO; Kessler et al. 1996). 
Unlike the all-sky IRAS mission, ISO could do pointed 
observations and offered multiple modes of operation. It 
gave us our first glimpse of the sky at wavelengths 
longer than 100~$\mu$m and revealed the richness of the 
infrared spectrum.

In this review, I will discuss the results obtained by ISO 
on pre-main sequence and Vega-type stars, focussing on the 
the evolution of circumstellar dust. In section~2 I will 
briefly introduce the capacities of the instruments on 
board ISO. The next section is devoted to a qualitative 
discussion of the infrared spectra of young stars, and 
outlines a possible evolutionary scenario for circumstellar 
dust. Section~4 deals with the originating region of 
the infrared radiation from pre-main sequence stars, 
whereas section~5 reviews the new picture of 
Vega-type stars that ISO had given us and their relation 
to young stars. In the last section I will 
summarize the main conclusions of this review and 
will look forward to the new questions that may be 
answered by future infrared missions.

\section{The Infrared Space Observatory}
ESA's {\it Infrared Space Observatory} consisted of a 60~cm 
cryogenically cooled telescope with four focal plane instruments, 
operating at wavelengths from 2.4 to 240 microns. Launched in 
November 1995, it collected more than 26,000 scientific observations 
before it finally ran out of liquid helium in April 1998, after 
a greatly successful mission that lasted eight months longer 
than expected.

The first of the four instruments, the 
infrared camera (ISOCAM; Cesarsky et al. 1996) covered the 
2.5--18~$\mu$m band with two different 32 $\times$ 32 pixel 
detectors. Each detector had several discrete band pass filters 
as well as circular variable filters (CVFs) with a resolution of about 
35. The second focal plane instrument, the photo-polarimeter (ISOPHOT; 
Lemke et al. 1996), covered the largest wavelength range of the ISO 
instruments: 2.5--240~$\mu$m. Its scientific capabilities included 
multi-filter and multi-aperture photometry, polarimetry, imaging and 
low-resolution ($R$ $\approx$ 100) spectrophotometry in the 2.5-4.9 and 
5.8-11.6~$\mu$m wavelength ranges.

The other two instruments on board ISO were two complementary spectrometers. 
The short-wavelength spectrometer (SWS; de Graauw et al. 1996) consisted 
of a scanning spectrograph with a spectral resolution ranging from 1200 
to 2400 covering the wavelength range of 2.4--45 $\mu$m. By inserting 
Fabry-P\'erot filters, the resolution of the instrument could be enhanced 
by a factor 20 in the wavelength range from 11.4 to 44.5~$\mu$m.
Similar to SWS, the ISO long-wavelength spectrometer (LWS; 
Clegg et al. 1996) consisted of a scanning spectrograph in the 
43--198~$\mu$m range, with a resolution of either 150--200, or 
6800--9700 in Fabry-P\'erot mode.

\section{Spectral evolution of pre-main sequence stars}
\subsection{Massive YSOs}
The ISO mission allowed us for the first time to explore those parts 
of the infrared spectrum that are hidden from view by the earth's 
atmosphere. Because of this, major steps forward have been made in 
our understanding of the star formation process.
In the standard picture of star formation (Shu et al. 
1987) four distinct phases can be distinguished. First, 
a number of dense cores develop within a molecular cloud. At some point, 
the dense cores collapse to form embedded protostars which accumulate 
material by accretion from the surrounding cloud. As the embedded 
protostars accrete mass, they also lose mass due to a strong stellar 
wind, which eventually leads to the dispersion of the reservoir of 
mass from the protostellar envelope and the end of accretion. The 
former protostar, now a pre-main sequence star surrounded by a 
circumstellar disk, begins to contract 
slowly, increasing its central temperature until hydrogen ignition 
takes place and it has become a stable star on the main sequence.

Spectra taken with all four focal plane instruments on board ISO have 
revealed a large variety of properties of gasses, dust and ices around 
YSOs that can be well interpreted in the scenario outlined above. 
The first stages of the spectral evolution of pre-main sequence 
objects are illustrated in Fig.~1, where we show ISO-SWS spectra of 
three massive YSOs, NGC~7538 IRS9 , AFGL 2591 and S106. The 
infrared spectrum of NGC~7538 IRS9 (Whittet et al. 1996) consists of 
a smooth continuum with broad, strong absorption bands due to silicates 
and ices. Frozen H$_2$O, CO, CO$_2$, CH$_4$, CH$_3$OH and a 
band around 6.8~$\mu$m whose carrier remains unidentified are 
all commonly detected in the lines of sight towards embedded 
YSOs and may be regarded as a strong indicator of youth.

The infrared spectrum of AFGL 2591 (van der Tak et al. 1999)
already shows 
striking differences with that of NGC~7538 IRS9. Although the 
massive young star is embedded to about the same degree as 
NGC 7538~IRS9, as evidenced by the similarity of the depth 
of their 10~$\mu$m silicate profile, the relative prominence of 
the ice absorption of some species, such as H$_2$O and CO$_2$ 
is diminished, whereas the absorption bands of CO, CH$_4$ and 
CH$_3$OH have completely vanished. In contrast to NGC~7538 IRS9, 
higher resolution ISO-SWS spectra of AFGL~2591 show a large 
number of sharp absorption lines due to gas-phase CO, CO$_2$ 
and H$_2$O (Helmich et al. 1996; van Dishoeck et al. 1996). 
The temperatures derived from these gaseous molecules 
amount to a few 100~K, well above the evaporation temperature 
of most ices.
\begin{figure*}[t]
\centerline{\psfig{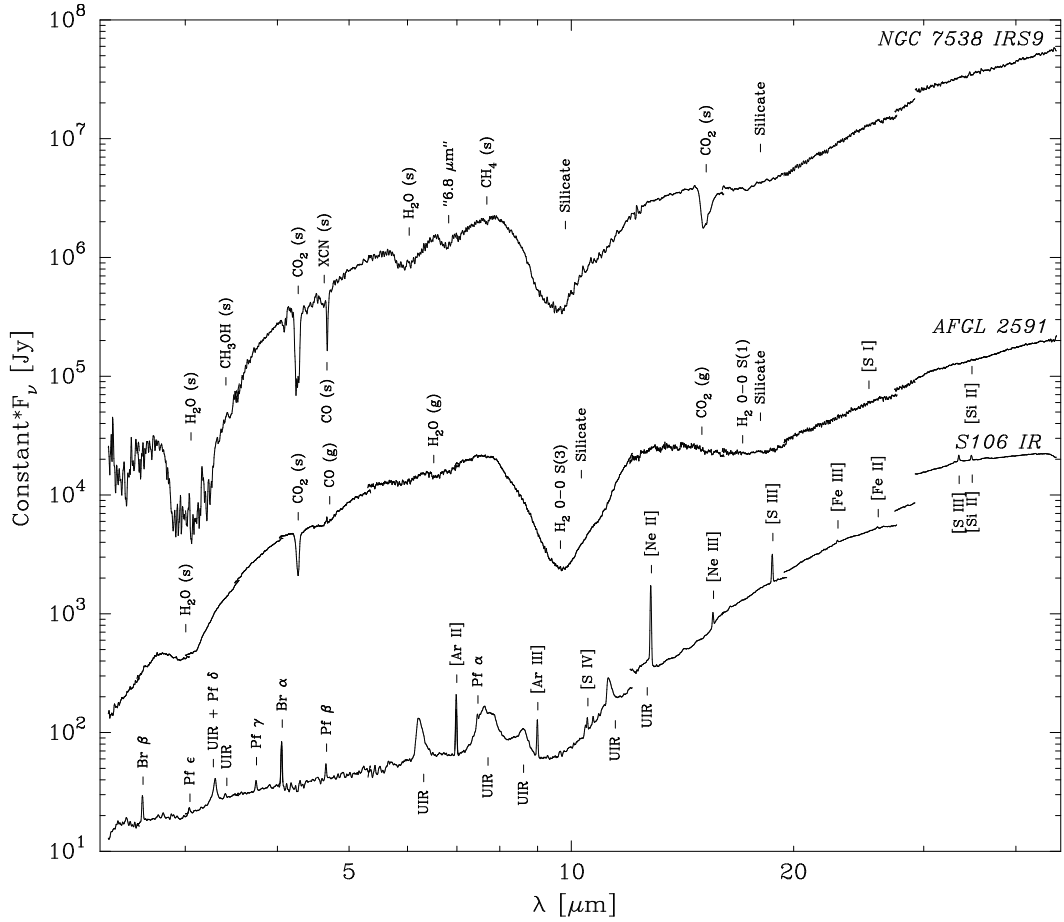}}
\caption[]{ISO-SWS spectra of the heavily embedded massive YSO 
NGC 7538 IRS9 (Whittet et al. 1996), the intermediately embedded massive YSO 
AFGL 2591 (van der Tak et al. 1999) and the bipolar nebula S106 
(van den Ancker et al. 2000a), 
illustrating the spectral evolution of high-mass YSOs.}
\end{figure*}
\begin{figure*}[t]
\centerline{\psfig{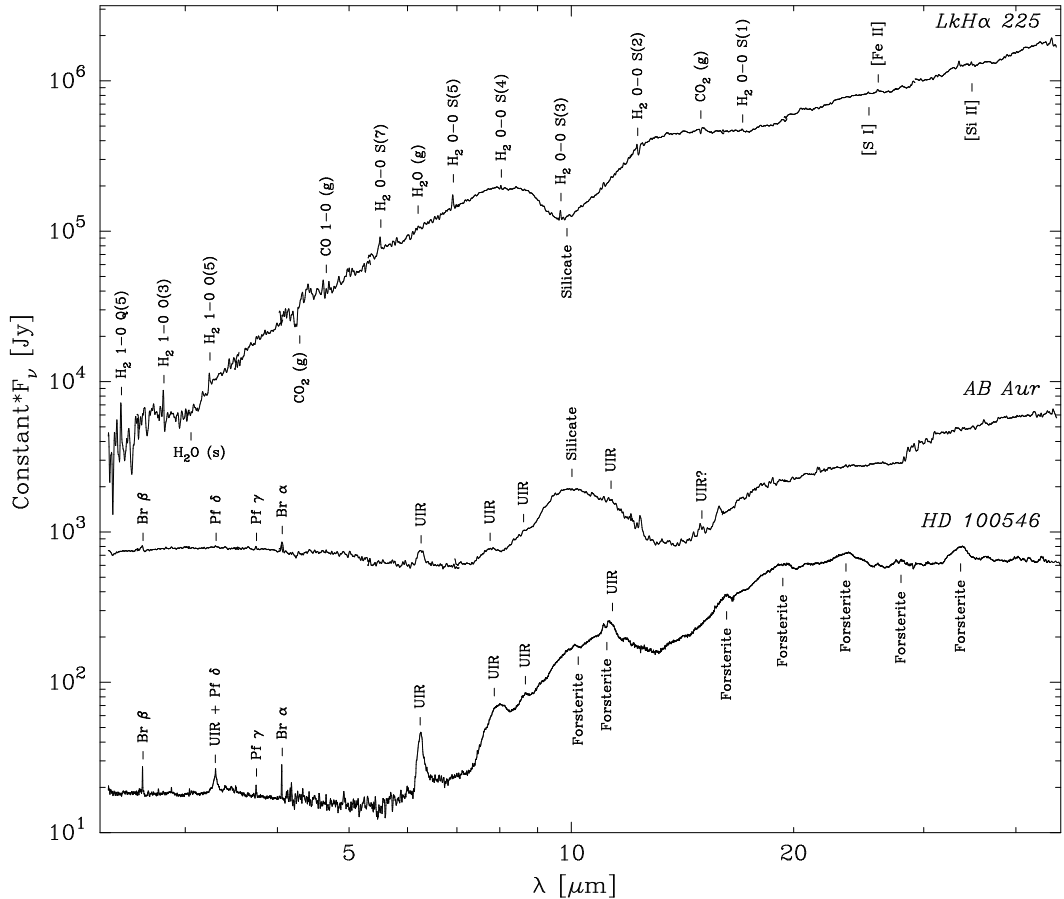}}
\caption[]{ISO-SWS spectra of the embedded intermediate-mass YSO 
LkH$\alpha$ 225 (van den Ancker et al. 2000b), and the Herbig Ae stars 
AB Aur (van den Ancker et al. 2000c) and HD~100546 (Malfait et al. 1998), 
illustrating the spectral evolution of intermediate-mass pre-main sequence 
stars.}
\end{figure*}

The difference between the infrared spectra of NGC~7538 IRS9 
and AFGL 2591 can easily be understood by {\it assuming} that 
the former is in an earlier evolutionary stage than the latter. 
Whereas the NGC~7538 IRS9 inner envelope consists of mostly 
unprocessed interstellar material, AFGL~2591 has already 
heated the inner region of its envelope to above the evaporation 
temperature of most ices and has formed a so-called ``hot core'' 
region. By analyzing gas/ice abundances in a larger sample of 
YSOs, van Dishoeck et al. (1999) found a good correlation 
between gas temperature and gas/ice abundance, showing that 
this process of the evaporation of ice mantles and the 
heating of the YSO's envelope is more general than illustrated 
here by the two examples of NGC~7538 IRS9 and AFGL~2591, 
and may in fact be a good indicator of age for embedded YSOs.

A second difference between the spectra of NGC~7538 IRS9 
and AFGL~2591 in Fig.~1 is the presence of emission lines 
due to H$_2$, [Si\,{\sc ii}] and [S\,{\sc i}] in AFGL~2591, 
which appear absent in NGC~7538 IRS9. In particular the 
detection of [S\,{\sc i}] 25.25~$\mu$m is significant, 
since this line can only be formed with detectable intensity 
in regions of shocked gas (van den Ancker et al. 2000b). This 
shocked gas is most likely the result of the interaction of 
AFGL~2591's outflow with the surrounding cloud and traces the 
clearing of the circumstellar envelope of a young star.

Compared to NGC~7538 IRS9 and AFGL~2591, the spectrum of S106 
(van den Ancker et al. 2000a), 
the third massive YSO shown in Fig.~1, looks very different. 
It consists of a smooth\footnote{The ``jumps'' seen at discrete 
wavelengths in the spectrum are due to the fact that the SWS 
uses entrance apertures of different sizes for different 
wavelength regions, so for a source that is not point-like, 
such as S106, one may receive more flux where such a change in 
aperture occurs.} continuum with a large number of sharp, 
strong emission lines due to recombination in the Brackett and 
Pfund series of H\,{\sc i}, and due to numerous fine-structure 
transitions of ionic species.
In addition to this, the unidentified infrared (UIR) emission 
bands, often attributed to polycyclic aromatic hydrocarbons 
(PAHs)\footnote{Note however that ISOCAM CVF images did not 
retrieve the dependence of the relative strength of the UIR 
bands on UV field expected for PAHs (Uchida et al. 2000 
and references therein).}, are present at 
3.3, 3.4, 6.2, 7.6, 7.8, 8.6, 11.3 and 12.7~$\mu$m. The 
central star in 
S106 has developed a strong stellar wind, as evidenced by 
the recombination lines, which has cleared most of the 
its natal cloud. The high ionization species in the spectrum 
show that it is now surrounded by a substantial H\,{\sc ii} 
region. The UIR emission as well as emission due to H$_2$ 
and other gas-phase molecules indicate that a so-called 
photo-dissociation region (PDR; see Hollenbach \& Tielens 
1999 for a comprehensive review), a mostly neutral region 
whose surface is photoelectrically heated under the 
influence of UV radiation, is present as well. The presence 
of this PDR emission shows that some regions of mostly 
neutral material, perhaps the remnants of the highest density 
clumps in the original cloud, remain.

\subsection{Intermediate-mass stars}
So far our discussion of YSO spectral evolution has focussed 
on massive objects. The infrared spectrum of these objects 
is typically dominated by their wide circumstellar environment 
and in fact evidence for the existence of protoplanetary disks 
around massive YSOs is 
weak. We now focus on intermediate-mass (2--10~M$_\odot$) 
pre-main sequence stars, the Herbig Ae/Be (HAeBe)
stars. In Fig.~2 we again show the ISO-SWS spectra of three 
of these, the embedded intermediate-mass YSO LkH$\alpha$ 225, 
and the two relatively isolated Herbig Ae/Be systems AB~Aur 
and HD~100546. 

Although there are some indications that deeply embedded 
intermediate-mass YSOs may have a different method of 
thermally processing their circumstellar ices (c.f the 
case of Elias~29, Boogert et al. 2000), the overall 
spectra of embedded intermediate-mass YSOs appear similar 
to those of high-mass YSOs. For example, the spectrum of 
LkH$\alpha$~225 (van den Ancker et al. 2000b) shown in 
Fig.~2 resembles that of many moderately embedded massive 
YSOs. Silicate absorption, 
as well as absorption due to water ice are visible and 
gas-phase CO, CO$_2$ and H$_2$O are present, indicative 
of a relatively evolved, warm core. Strong shock-excited 
emission lines of H$_2$, [S\,{\sc i}], [Fe\,{\sc ii}] 
and [Si\,{\sc ii}] are present, again most likely 
related to the presence of an outflow from this star.

However, the next stage in the evolution of a intermediate-mass 
YSO, represented here by the well-studied Herbig Ae star AB~Aur 
(van den Ancker et al. 2000c), 
yields a very different spectrum than that in the corresponding 
stage in the life of a massive star. The lines from the Pfund 
and Brackett series of H\,{\sc i}, indicative of a strong wind, 
are still present, but the spectrum lacks the multitude of 
atomic fine-structure lines seen in the spectrum of S106, 
indicating that the central star of AB~Aur does not emit 
sufficient UV photons to create a substantial H\,{\sc ii} region. 

Several UIR bands can be seen in the spectra of 
AB~Aur and HD~100546 shown in Fig.~2. 
An analysis of all spectra of Herbig Ae/Be stars found in 
the ISO data archive shows that the UIR bands are present 
in about 50\%, without a strong preference for a certain 
spectral type. Whereas undoubtedly in some cases the 
observed UIR emission arises in filamentary PDR 
structures included in the rather large beam of the 
ISO spectrometers (e.g. 20\arcsec $\times$ 14\arcsec\ for 
the SWS), ground-based imaging as well as ISOCAM CVF data 
show that more commonly the UIR emission arises close 
to the star. At present it is not clear why some Herbig 
Ae/Be stars show this compact UIR emission, whereas 
others with apparently very similar properties do not. 
However, the absence of UIR bands in some Herbig 
Ae/Be stars does show that models depending on the 
presence of very small dust grains to explain the 
observed near-infrared excess in Herbig Ae/Be stars 
%(Natta et al. 1993; Natta \& Kr\"ugel 1995) 
will not be successful in all cases.

The answer to the question why only some HAeBes show 
UIR emission may come from the observation that a few 
objects (e.g. AB~Aur, Fig.~2) exhibit strong UIR bands 
due to C--C stretching  modes, whereas the bands due to 
C--H bonds are weak or absent, suggesting that either 
the PAHs have been stripped of most of their hydrogen 
atoms, or that they are rather large ($>$ 100 C atoms). 
The presence of the rare UIR bands at 3.43
and 3.53~$\mu$m in a small number of Herbig stars (most 
noticeably HD~97048 and Elias~1; Van Kerckhoven et al. 1999) 
may reflect an even further chemical processing stage, 
in which nanodiamonds, not unlike those found in 
meteorites in our own solar system, have been 
formed (Guillois et al. 1999).

Although C-rich dust is certainly present, 
the general appearance of the infrared spectrum of most 
Herbig Ae/Be stars is determined by O-rich dust particles. 
Broad emission features in the 8--12 and 17--19~$\mu$m 
ranges due to silicate O--Si--O bending and stretching modes 
are found in a large fraction of Herbig Ae/Be stars. 
In some more embedded objects they are seen in absorption. 
At longer wavelengths, emissions due to water ice and
hydrous silicates have been reported in some Herbig 
stars (Malfait et al. 1998, 1999). A broad spectral 
feature ranging from 14 to 38~$\mu$m was observed by 
van den Ancker et al. (2000c) in the Herbig Ae/Be stars 
AB~Aur and HD~163296, which could due to a blend of the 
28~$\mu$m resonance of iron-oxide with the 19~$\mu$m 
resonance of amorphous silicates. Interestingly, 
iron or iron-oxide at $\sim$ 1000~K can also naturally 
explain the near-infrared excess observed in virtually 
all HAeBes.

The fact that the silicate features are seen in emission 
suggests that the emitting material is optically 
thin at 10~$\mu$m in the objects where this is the case. 
This puts strong constraints on the distribution of dust 
and its temperature structure. However, millimeter continuum 
observations imply that the masses contained in these 
disks can be substantial, leading some authors  to suggest 
that the observed silicate emission could arise in an 
extended optically thin region above the dense mid-plane 
of the disk (Natta et al. 1999; Bouwman et al. 2000.

In most stars the 9.7~$\mu$m silicate emission appears broad 
and smooth, such as in AB~Aur. It is essentially the inverse 
of the very smooth silicate features seen in absorption towards 
all embedded YSOs (e.g. AFGL~2591 and LkH$\alpha$~225 
in Figs.~1 and 2). Comparison with laboratory determinations 
of optical constants of silicates shows that the lattice 
structure of these silicates must be amorphous, i.e. 
relatively unordered. Although most common, not all Herbig
stars show the very smooth silicate emission profiles 
seen in AB~Aur. An especially spectacular example is that 
of the Herbig B9 star HD~100546 (Waelkens et al. 1996; 
Malfait et al. 1998), shown as the bottom curve in Fig.~2. 
Here we see that the 8--12~$\mu$m silicate feature shows 
a lot of sub-structure, and many broad bands with strengths 
up to that of the underlying continuum appear in 
emission at the longer wavelengths of the SWS wavelength 
range. These sharper features are indicative of silicates 
in which the lattice structure is more crystalline. 
Note that detailed modelling of the dust composition  
shows that even though the crystalline silicates dominate 
the appearance of the infrared spectrum, the bulk of 
the silicates is still in the amorphous form 
(Malfait et al. 1999). Comparison with lab data shows that 
the peaks seen in the HD~100546 spectrum must be due to 
the mineral forsterite (Mg$_2$SiO$_4$). 
This is exactly the same material as is found in comets in our 
own solar system (Crovisier et al. 1997), suggesting that the 
processing of circumstellar dust in the primitive solar system 
bears similarity to that in HAeBes.

\subsection{T Tauri stars}
The SWS and LWS were sufficiently sensitive to allow the 
study of massive YSOs at distances up to several kiloparsecs, 
as well as the nearest intermediate-mass stars, but they 
lacked the sensitivity to study even the nearest T~Tauri 
stars, low-mass pre-main sequence stars which may represent a 
better analog of the protosolar nebula than the Herbig Ae/Be 
stars. However, the spectroscopic modes of the more 
sensitive ISOPHOT and ISOCAM instruments, although limited 
to wavelengths smaller than 18~$\mu$m, also allowed valuable 
information on the dust composition in the nearest young 
stars of solar mass to be gathered. Based on ISOPHOT data, 
Natta et al. (2000) show that the amorphous silicate 
emission features exhibited by most Herbig Ae stars are also 
found around all nine studied T~Tauri stars in the 
Chamaeleon~I dark cloud. Possibly due to the higher 
requirements with respect to signal to noise and spectral 
resolution, crystalline silicates still await their 
first detection in a low-mass YSO. Although a detailed study 
of the dust composition of T~Tauri objects awaits more sensitive 
mid-IR spectroscopic instruments such as the infrared spectrograph 
on board NASA's {\it Space Infrared Telescope Facility} (SIRTF), 
there are at the time of writing no indications that the 
spectral evolution sketched in the previous subsection for 
intermediate-mass YSOs would not equally apply to solar-mass 
pre-main sequence stars.

\section{Disks and envelopes around young stars}
The IRAS mission detected infrared radiation, believed 
to originate in circumstellar disks or envelopes, from many 
young stellar objects. However, its poor spatial resolution 
inhibited a comprehensive study of young clusters. 
Higher spatial resolution at wavelengths not hindered by 
interstellar extinction allowed ISO's camera to provided a 
new complete census of nearby star forming regions, 
resulting in a doubling of the known number of YSOs 
in each (Olofsson et al. 1999). The new cluster members 
are mainly at the low-luminosity end of the luminosity 
distribution, indicating that the initial mass function 
continues to rise slowly (with index $-$0.2) until brown 
dwarf masses.

Preliminary results 
of a ISOPHOT study of 97 T~Tauri stars in five young clusters 
have been reported by Robberto et al. (1999). They found the  
fraction of stars with 60~$\mu$m excesses to decrease from 
around 60--70\% in Taurus, $\rho$~Oph and the R~CrA region 
to below 20\% in the older Cham~I and TW~Hydra associations. 
More than 80\% of the stars in these clusters that 
exhibit near-infrared excess emission also show a far-infrared 
excesses. In contrast, only one star that has no near-infrared 
excess has one at 60~$\mu$m. No 60~$\mu$m excesses were 
detected in the eldest clusters studied by Robberto et al., 
suggesting that for low-mass stars the dust in disks between 
0.3 and 3~AU disappears on timescales of $\sim$ 10~Myr.

A long-standing problem in studies of pre-main 
sequence stars has been the question whether the observed 
infrared excesses arise in a relatively stable circumstellar 
disk or in an infalling envelope. This question can be 
answered by studying the spatial extent of sources in the 
infrared. \'Abrah\'am et al. (2000) observed seven Herbig 
Ae/Be stars at mid-infrared wavelengths with ISOPHOT and 
concluded that at $\lambda$ $<$ 25~$\mu$m the observed 
emission mainly arises from a compact area, whereas at 
longer wavelengths it is spatially extended. At wavelengths 
longer than 100~$\mu$m, the emission is never dominated 
by the Herbig Ae/Be stars, but by dust cores of about 
1\arcmin\ in size. These results show that at least 
in these cases the observations exclude a ``disk-only'' 
geometry and a (not necessarily spherical) envelope 
needs to be present as well. They do not exclude the 
possibility that in some cases only a disk remains. 
Further study of a sample that also contains Herbig stars 
located in relative isolation is required to provide 
the definitive answer to this problem.

Although the infrared continuum will be dominated by the 
circumstellar dust, the gas in the disk will also produce 
spectral signatures in the infrared. The pure rotational 
lines of H$_2$, all located in the mid-infrared, are 
especially interesting since they directly probe the 
dominant gas component in the disk and hence allow us 
to measure disk masses without relying on an essentially 
unknown CO/H$_2$ ratio. For most YSOs observed with 
ISO the observed H$_2$ appears too warm (typically 500~K) 
to be associated with the circumstellar disk: it probably 
arises in the loose stellar surroundings and is associated 
with either PDRs or with shocked gas in outflows (e.g. 
van den Ancker et al. 2000a, b). However, van Dishoeck et al. 
(1998) report the detection of the H$_2$ 0--0 S(0) and S(1)
lines toward the isolated Herbig Ae star HD~163296. The 
implied temperature of only 150~K, the disappearance of 
the H$_2$ emission off-source and the absence of 
ro-vibrational H$_2$ emission in ground-based imaging 
all argue in favor of an interpretation in which the 
H$_2$ lines observed in HD~163296 originate in the 
circumstellar disk. The total observed gas mass in 
this object is 0.007~M$_\odot$. 
The value of 35 for the gas to dust ratio in the 
HD~163296 disk that is derived by combining this gas 
mass with the dust mass in the HD~163296 computed 
by Bouwman et al. (2000) is surprisingly close to 
the canonical value of 100 for the gas/dust ratio in 
the interstellar medium.

\section{Vega-type Stars}
One of the most unexpected results of the IRAS mission was 
the discovery of an infrared excess due to the presence of 
circumstellar dust around several nearby main-sequence stars 
(Aumann et al. 1984). Spectacular images showing an edge-on disk 
around the relatively young main-sequence star $\beta$~Pictoris 
(Smith \& Terrile 1984) raised the question whether these disks 
are a remnant of the star formation process, or represent the 
extrasolar equivalents of the Kuiper belt in our own solar system. 
However, the IRAS survey and subsequent ground-based follow-up 
research lacked the sensitivity to determine what fraction of 
nearby stars show the Vega phenomenon. 

Using the ISO photometer, Dominik et al. (1999) performed a deep 
survey on a volume-limited sample of nearby main-sequence stars 
of spectral types A, F, G and K. They found infrared excesses 
in approximately 21\% of nearby stars. However, Vega-type phenomena 
appear much more common in A- and F-type than in stars of later 
spectral type. This need not necessarily reflect a different 
disk mass, but may be due to a mixture of disk illumination and 
time evolution. Based on the same data-set, Habing et al. (1999) 
report a distinctly bimodal 
distribution of the frequency of dust disks with age: more than 
50\% of stars younger than 300~Myr have infrared excesses at 
25, 60 or 170~$\mu$m, while most stars older than 400~Myr do not. 
Ninety percent of the disks disappear when the star is between 
300 and 400~Myr old. These timescales are similar to those 
derived for the final-clean up phase of our own solar system. 
This result clearly shows that in most cases the Vega-type 
phenomenon represents remnants of the star formation process 
and that the vicinity of the sun hosts a considerable 
population of relatively young stars.

However, not {\it all} debris disks decay after 400~Myr. In about 
10 percent of the cases circumstellar dust persists for longer 
than 400~Myr. At present is is unclear what causes these disks 
to live longer than most. A hint to the answer may come from 
the detection of a 60~$\mu$m excess in $\rho^1$ Cnc 
(Dominik et al. 1998). Around this 4~Gyr-old star both 
a planet and a disk are present and it is not inconceivable 
that the presence of a third body may be required to keep 
a circumstellar disk stable on long time scales.

Soon after their discovery it was realized that the timescale for the 
removal of dust particles around Vega-type stars is much smaller 
than their age: radiation pressure will push the small dust particles 
out of the system, whereas Poynting-Robertson drag will bring the 
larger ones in to where they are evaporated. Therefore they must 
be re-supplied, at least for some fraction of the stellar lifetime.
Possibilities that have been speculated upon to provide the 
mechanism for the replenishment of small dust grains include 
collisions between larger bodies or the evaporation of 
extra-solar comets.

Indications to the origin of the dust seen in the 
Vega-type systems may come from infrared spectroscopy. Solar 
system comets show a number of distinctive spectral signatures 
in the infrared due to the presence of crystalline silicate 
dust (e.g. Crovisier et al. 1997). Ground-based spectroscopy 
of the $\beta$~Pic system in the 10~$\mu$m atmospheric 
window seemed to suggest a high degree of similarity of 
the dust seen in this system to that in comets (Telesco \& 
Knacke 1991; Knacke et al. 1993). 
However, Pantin et al. (1999) report an absence of the 
expected mid-infrared crystalline silicate features in 
the ISO-SWS spectra they obtained of $\beta$~Pic. At 
present it is unclear whether comets need to be 
discarded as the source of dust in Vega-type systems, 
or that the in many ways unusual $\beta$~Pic disk 
has a different composition than most Vega-type stars. 
Infrared spectroscopy of fainter, more typical, 
Vega-type stars with SIRTF will be able to clarify 
this issue within the coming years.

\section{Concluding remarks}
The {\it Infrared Space Observatory} has yielded 
great new insights in the the evolution of disks into 
planetesimals and planets. For the 
first time we are able to analyze the chemical 
composition and evolution of circumstellar dust 
and only after the ISO mission has it been 
realized how important solid-state resonances are 
for the interpretation of infrared energy 
distributions. In fact the flux in all four IRAS 
bands, centered at 12, 25, 60 and 100~$\mu$m, can 
be dominated by solid-state emission features, 
demonstrating the necessity of infrared spectroscopy
to interpret the energy distributions of pre-main 
sequence objects.

One of the most exciting ISO results has been the 
discovery of crystalline silicates, such as found in 
our own solar system, in some Herbig Ae/Be stars. 
The fact that in embedded YSOs the silicates are 
invariably amorphous shows that for intermediate-mass 
stars, the transition from amorphous to crystalline 
must occur in the HAeBe phase of evolution. The 
mechanism responsible for this transition is yet 
unclear. Laboratory experiments show that crystallization 
of silicates may be achieved by either a slow annealing 
of warm grains, or by heating the dust grains to 
temperatures above $\sim$ 1000~K. Since the crystalline 
dust in Herbig Ae/Be stars is observed to have much 
lower temperatures, substantial mixing in the disk 
would be required in this scenario. However, a study 
of ISO spectra of post-AGB stars by Molster et al. (1999)
shows that in these objects crystallization occurs 
at much lower temperatures over long time-scales. They 
suggest that the same low-temperature annealing 
process at work in these stars may also be responsible 
for the crystallization of silicates in young stars.
If proven, this same process will be responsible for 
the creation of crystalline silicates in solar-system 
comets, a conclusion which ultimately may alter our 
view of the origin of our own solar system.

ISO has also left us with a feeling for the richness 
and variety of the infrared spectra of young stars, 
resulting in the qualitative picture of spectral 
evolution sketched in section~3. However, a 
quantitative conformation of this picture awaits 
bigger samples and in particular its extension 
towards solar-mass young stars. 
Indeed, van den Ancker et al. (2000c) remark upon 
the fact that stellar ages as derived from the position 
in the Hertzsprung-Russell diagram do not show a 
one-to-one correlation with degree of crystallinity. 
The fact that in a typical young cluster only half of 
the pre-main sequence objects shows an infrared excess 
(e.g. Beckwith et al. 1990), 
suggests that this may be due to greatly differing 
time-scales for disk evolution from object to object. 
The lack of correlation between submm spectral index, 
an indicator for grain growth, and degree of 
crystallinity noted by these authors is more puzzling. 
Apparently these two processes are not necessarily 
coupled, suggesting that processes other than 
stellar mass and age, perhaps the environment of 
the star, are important.

The discovery that the Vega-type phenomenon is common 
in relatively young main-sequence stars, whereas it is 
rare in stars older than 400~Myrs has considerably 
strengthened the connection between pre-main sequence 
and Vega-type stars. However, the final proof of 
an evolutionary link between these two categories 
of objects still remains to be made and most 
likely will be made soon: the last of 
NASA's great observatories, SIRTF, is scheduled 
to be launched in the near future and, with 
its greatly enhanced sensitivity, promises to 
continue the path of discoveries that has been 
explored by ISO.

\end{document}